\documentclass[12pt,letter]{iopart}

\usepackage{graphicx,amssymb}

\begin{document}
 
\title{Covalent bonding and the nature
 of band gaps in some half-Heusler compounds}

\author{Hem Chandra Kandpal\dag, Claudia Felser\dag,\\ and Ram Seshadri\ddag}

\address{\ddag Institut f\"ur Anorganische Chemie und Analytische Chemie\\
Johannes Gutenberg-Universit\"at, Staudinger Weg 9, 55099 Mainz\\
         felser@uni-mainz.de}

\address{\ddag Materials Department and Materials Research Laboratory\\
         University of California, Santa Barbara CA 93106\\
         seshadri@mrl.ucsb.edu}

\begin{abstract}

Half-Heusler compounds \textit{XYZ}, also called semi-Heusler compounds,
crystallize in the MgAgAs structure, in the space group $F\bar43m$. We report 
a systematic examination of band gaps and the nature (covalent or ionic) 
of bonding in semiconducting 8- and 18- electron half-Heusler compounds
through first-principles density functional calculations. We find 
the most appropriate description of these compounds from the viewpoint of 
electronic structures is one of a \textit{YZ} zinc blende lattice stuffed  
by the \textit{X} ion. Simple valence rules are obeyed for bonding
in the 8-electron compound. For example, LiMgN can be written
Li$^+$ + (MgN)$^-$, and (MgN)$^-$, which is isoelectronic with (SiSi),
forms a zinc blende lattice. The 18-electron compounds can similarly be
considered as obeying valence rules. A semiconductor
such as TiCoSb can be written Ti$^{4+}$ + (CoSb)$^{4-}$; the latter unit
is isoelectronic and isostructural with zinc-blende GaSb. For both the
8- and 18-electron compounds, when \textit{X} is fixed as some electropositive
cation, the computed band gap varies approximately as the difference in Pauling 
electronegativities of \textit{Y} and \textit{Z}. What is particularly 
exciting is that this simple idea of a covalently bonded \textit{YZ} lattice 
can also be extended to the very important \textit{magnetic\/} half-Heusler 
phases; we describe these as valence compounds \textit{ie.\/} possessing a 
band gap at the Fermi energy albeit only in one spin direction. The 
\textit{local\/} moment in these magnetic compounds resides on the 
\textit{X} site.

\end{abstract}


\maketitle

\section{Introduction}

The half-Heusler phases \textit{XYZ}, comprising three interpenetrating 
\textit{fcc} lattices, constitute an important class of materials with 
particular regard to their magnetic properties. de Groot\cite{deGroot}
and coworkers showed a number of years ago that the half-Heusler compound
NiMnSb can be described as a half-metallic ferromagnet, whose computed band 
structure resembles a metal in one spin direction, and a 
semiconductor in the other. Since then, and indeed, even prior to that, it has
been recognized that the electronic structure and hence properties of Heusler 
compounds\cite{Kubler} and half-Heusler compounds\cite{Pierre1,Pierre2} are 
very sensitive to the valence electron count. 

A number of electronic structural studies have been carried out on the 
half-Heuslers. We focus here on those studies which systematically address
behavior in families of half-Heusler compounds, rather than studies focused
on individual ones. From the viewpoint of chemical bonding in these compounds,
Whangbo and coworkers\cite{Whangbo} have examined the non-magnetic band 
structures, using the extended H\"uckel method, of a number of half-Heusler 
compounds with varying valence electron counts. These authors have recognized 
that many \textit{XYZ} half-Heuslers can be thought of as comprising an 
\textit{X}$^{n+}$ ion stuffing a zinc blende \textit{YZ}$^{n-}$ sublattice
where the number of valence electrons associated with \textit{YZ}$^{n-}$ are
18 ($d^{10}$ + $s^2$ + $p^6$). 18 electron compounds are therefore closed shell
species; non-magnetic and semiconducting. They further suggest that the 
17 and 19 electron \textit{XYZ} would undergo a Stoner 
instability\cite{Stoner} to a ferromagnetic ground state, while the 22
electron compounds (typically with Mn$^{3+}$ at the \textit{X} site) should
be localized moment ferromagnets. The 22 electrons divide themselves
into 13 in the majority spin and 9 in the minority spin direction, resulting in
a semiconducting gap (half-metallic behavior) in the minority spin direction. 
Recently, Galanakis \textit{et al.\/}\cite{Galanakis} have placed this ``18 
electron'' rule on a more formal footing. 

Pierre \textit{et al.\/}\cite{Pierre1} were amongst the first to recognize the 
importance of the valence electron count in the half-Heuslers. In more recent 
work, Tobola and Pierre\cite{Pierre2} have emphasized the importance of 
covalency in these compounds. The \textit{Z} element is often a pnictogen 
(As, Sb or Bi) or some other main group element because only covalent bonding 
would justify the somewhat open half-Heusler structure. \"O\u{g}\"ut and 
Rabe\cite{Ogut} have examined the electronic structures of the compounds 
\textit{X}NiSn with \textit{X} = Ti, Zr, or Hf, and interpreted phase 
stability and the nature of band gaps. Also important from the approach that 
we will take here is the work of Wood, Zunger, and de Groot\cite{Wood} who have 
examined a number of non-magnetic ``stuffed zinc blende'' semiconductors, 
the so-called Nowotny-Juza compounds\cite{Nowotny,Juza} including 
half-Heusler phases such as LiZnP, in order to control the nature of the band 
gap (direct or indirect). Nanda and Dasgupta\cite{Dasgupta} have examined 
nearly 20 different half-Heusler compounds using the FP-LMTO and LMTO-ASA 
methods, including a detailed analysis of the bonding and the nature
of the band gaps. They argue as we do here, for the very important role played
by covalent bonding in these systems. They ascribe half-metallicity to arise
in some of the half-Heuslers due to the large \textit{Y}-\textit{Z} covalency,
in conjunction with large exchange -splitting due to highly magnetic \textit{X}
ions. 

In this contribution, we focus on chemical bonding in 8, 18 and magnetic
half-Heusler compounds, and attempt to relate the electronic structure 
to simple concepts such as the electronegativity of the component species.
We also demonstrate covalency and the local nature of the magnetic moment
in the magnetic compounds using real-space descriptors derived from first 
principles theory.

\section{Crystal structures and methods}

\begin{figure}
\centering \includegraphics[width=6cm]{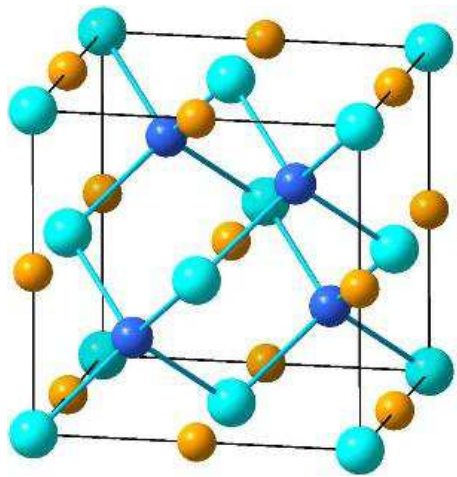}
\caption{(Color) \textit{XYZ} Half-Heusler crystal structure in the 
$F\bar43m$ space group. Cyan \textit{Z} atoms are at the origin, orange 
\textit{X} at $(\frac12, \frac12, \frac12)$ and blue \textit{Y} at 
$(\frac14, \frac14, \frac14)$. Note the tetrahedral zinc blende (diamondoid) 
sublattice formed by \textit{Y} and \textit{Z}.}
\end{figure}

Half-Heusler \textit{XYZ} compounds crystallize in the space group of zinc 
blende ($F\bar43m$) with a cubic cell parameter near 6.0\,\AA. The least and 
most electronegative elements are \textit{X} at $(\frac12,\frac12,\frac12)$ 
and \textit{Z} at $(0,0,0)$ forming a rock salt lattice. \textit{Y} are found
at $(\frac14,\frac14,\frac14)$ in the centers of tetrahedra formed 
by \textit{Z}, as well as by \textit{X}. Connecting \textit{Y} and \textit{Z} 
reveals the stuffed zinc blende lattice of the half-Heusler structure displayed
in Figure\,1. There are other, equivalent descriptions of this structure, but 
this is the one we chose, because it is closest to our description of the 
chemical bonding.

Density functional theory-based electronic structure calculations were 
performed using the full-potential Linear Augmented Plane Wave (LAPW) code 
\textsc{Wien2k}\cite{Wien} to optimize cell volumes of the different 
half-Heusler compounds described here. The electronic structural descriptions 
made use of Linear Muffin Tin Orbital (LMTO) calculations within the local spin
density approximation, as implemented in the \textsc{Stuttgart tb-lmto-asa} 
program.\cite{StuttgartLMTO} Starting structures for LMTO calculations were 
obtained from the results of volume optimization using \textsc{Wien2k}. Two 
important tools have been used to visualize the electronic Structure of these 
phases. The crystal orbital Hamiltonian population (COHP) \cite{COHP} enables 
the repartitioning of densities of states into regions which are pairwise 
bonding, non-bonding, and antibonding. The electron localization function 
(ELF)\cite{Becke,Silvi} is a real-space indicator of the extent to which 
electrons are localized, and display a strong Pauli repulsion. The ELF 
therefore serves to locate bonding and non-bonding electron pairs in the real 
space of the crystal structure. A real-space bonding analysis of half-Heusler 
compounds has not, to our knowledge, been previously carried out.

\section{Results and discussion}

\subsection{8-electron compounds}

\begin{table}
\centering \begin{tabular}{|l|l|l|l|l|l|l|l|}
\hline
\multicolumn{8}{|c|}{8 electron compounds with \textit{X} = Li}\\
\hline \hline
\textit{XYZ} & $a_{\mathrm{Calc.}}$ (\AA) &  
       $a_{\mathrm{Exp. }}$ (\AA) & $\chi_X$ &  $\chi_Y$ 
                &  $\chi_Z$ & Gap (eV) & $B$ (GPa) \\
\hline
LiMgN   & 5.072 & 4.955  & 0.98  & 1.31 & 3.04 &  2.51 & 80.1 \\  
\hline
LiMgP   & 6.028 & 6.021 & 0.98  & 1.31 & 2.19 &  1.92 & 49.6 \\  
\hline
LiMgAs  & 6.218 & 6.19  & 0.98  & 1.31 & 2.18 &  1.55 & 42.9 \\  
\hline
LiMgBi  & 6.803 & 6.74  & 0.98  & 1.31 & 2.02 &  0.64 & 30.3 \\  
\hline
LiZnP   & 5.707 & 5.779 & 0.98  & 1.65 & 2.19 &  1.23 & 65.4 \\  
\hline
LiCdP   & 6.118 & 6.087 & 0.98  & 1.69 & 2.19 &  0.85 & 52.8 \\  
\hline
LiAlSi  & 5.937 & 5.930 & 0.98  & 1.61 & 1.90 &  0.45 & 62.8 \\  
\hline
\end{tabular}
\caption{Results of density functional calculations on Li\textit{YZ} phases,
with experimental cell parameters for comparison. Cell parameters and 
bulk moduli are from LAPW calculations, and band gaps from LMTO calculations.}
\end{table}

We have examined 7 compounds Li\textit{YZ} with the half-Heusler structure
using LAPW and LMTO calculations. Results from the calculations
are summarized in Table\,1.  The calculated cell parameters match very well
with experimental cell parameters obtained from standard 
tabulations.\cite{Pearsons} For LiMgN, the experimental cell parameter
is from Kuriyama \textit{et al.}\cite{kuriyama}
This suggests that in all cases, the assignment 
of atomic positions, which is not always a simple matter to determine from
x-ray diffraction, is well justified.  We discuss these results here in detail.

\begin{figure}
\centering \includegraphics[width=11cm]{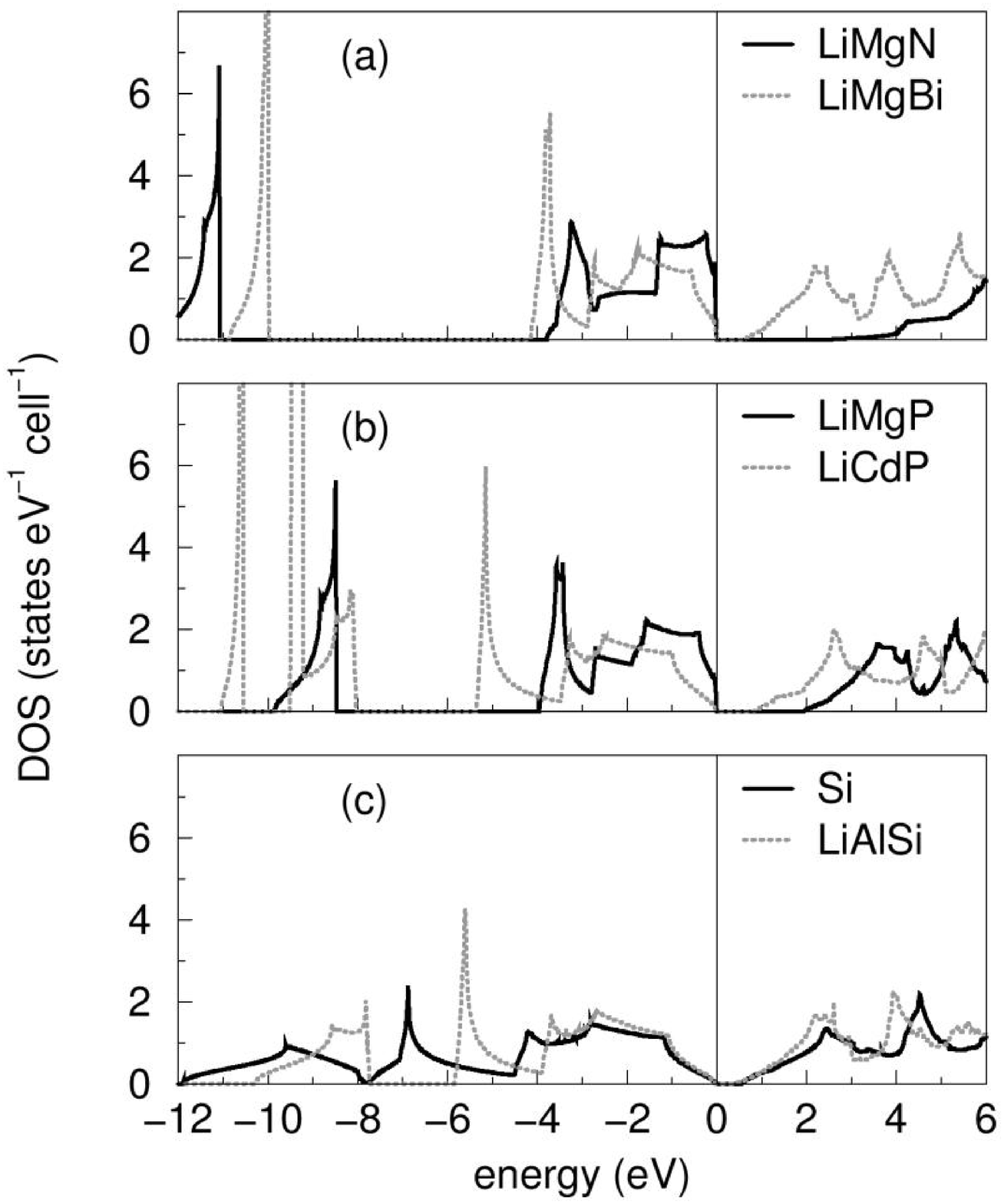}
\caption{LMTO densities of state for a number of different Li-based
half-Heusler phases compared with diamond Si.}
\end{figure}

In the three panels of Figure\,2, we compare the densities of states near the
Fermi energy (taken as the top of the valence band, and set as the origin)
for 6 different Li\textit{YZ} half-Heusler compounds. In panel (a) of this
figure two compounds LiMgN and LiMgBi are compared. Both compounds have a well
defined band gap. Replacing N with the heavier pnictogens P, As, (DOS
not shown) or Bi found to narrow the band gap due to the increasing
band width of both the valence and conduction bands. If the \textit{X} and 
\textit{Z} ions are held constant, as in LiMgP and LiCdP, we observe that 
replacing the more ionic Mg by the softer Cd also results in a narrowing of the
band gap. In panel (c) we compare LiAlSi with isoelectronic Si in the diamond 
structure. The electronic structures display a remarkable similarity in the
nature and extents of the valence and conduction bands. This strong similarity
has been noted previously by Christenson\cite{Christenson} and is fully
in keeping with the description of LiAlSi being a Zintl or valence 
compound,\cite{Zintl} wherein the identical electron counts of (AlSi)$^-$ and 
(SiSi) in turn imply that the structures would be similar. It should be noted
that Mg$_2$Si in the fluorite structure also obeys the same rule, if
we recognize it can be recast as Mg$^{2+}$(MgSi)$^{2-}$ with the (MgSi)$^{2-}$ 
crystallizing in a zinc blende lattice. Indeed the electronic structure of 
MgSi$_2$\cite{Froseth} is quite similar to what we find for LiAlSi. 
The change in the space group, comparing ($Fm\bar3m$) Mg$_2$Si and 
($F\bar43m$) LiAlSi arises because the atoms in the \textit{X} and \textit{Y} 
sites in the former are identical.

\begin{figure}
\centering \includegraphics[width=11cm]{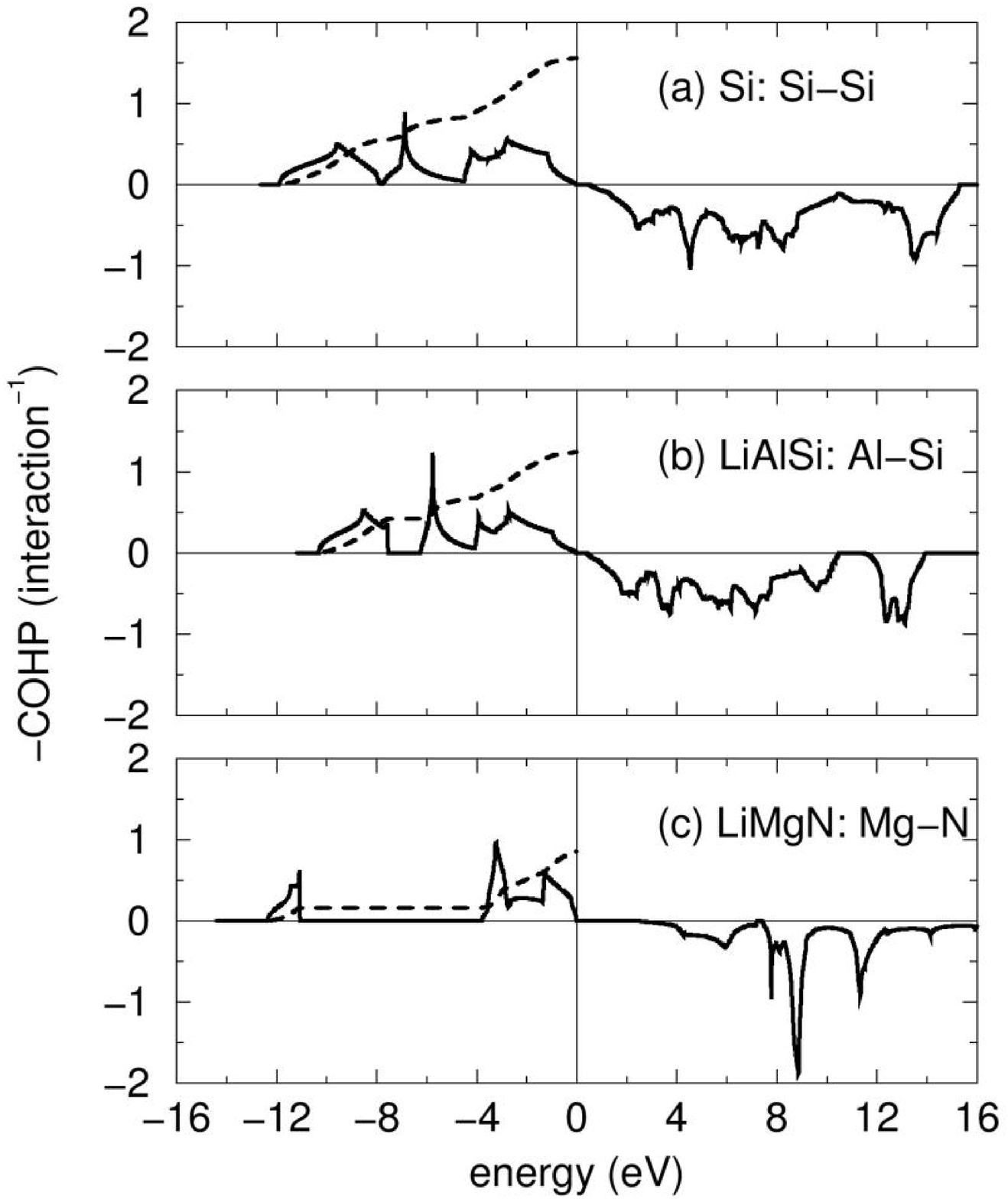}
\caption{Crystal orbital Hamiltonian populations (COHPs) for pairwise 
interactions in diamond Si, LiAlSi, and LiMgN. The dashed lines are 
integrations of the COHPs.}
\end{figure}

The similarity in the electronic structures of Si and stuffed
zinc blende compounds is further emphasized through an analysis of the Si-Si,
the Al-Si, and the Mg-N COHPs of the three different compounds: Si, LiAlSi, 
and LiMgN, shown in Figure\,3. The dashed line in this figure is an integration 
of the COHP up to $E_{\rm F}$, yielding a number that is indicative of the 
strength of the bonding. Not only are the extents of the bonding and 
antibonding COHPs of Si and LiAlSi very similar, but so is the value of
the integrated COHP: near -1.5\,eV \textit{per\/} interaction for Si and 
for LiAlSi. We interpret this as indicative of very similar extents of 
covalency in the diamond lattice of Si and the zinc blende sublattice of
LiAlSi. This value is slightly reduced in the more polar LiMgN, and the nature 
of the COHP is different as well. 

\begin{figure}
\centering \includegraphics[width=12cm]{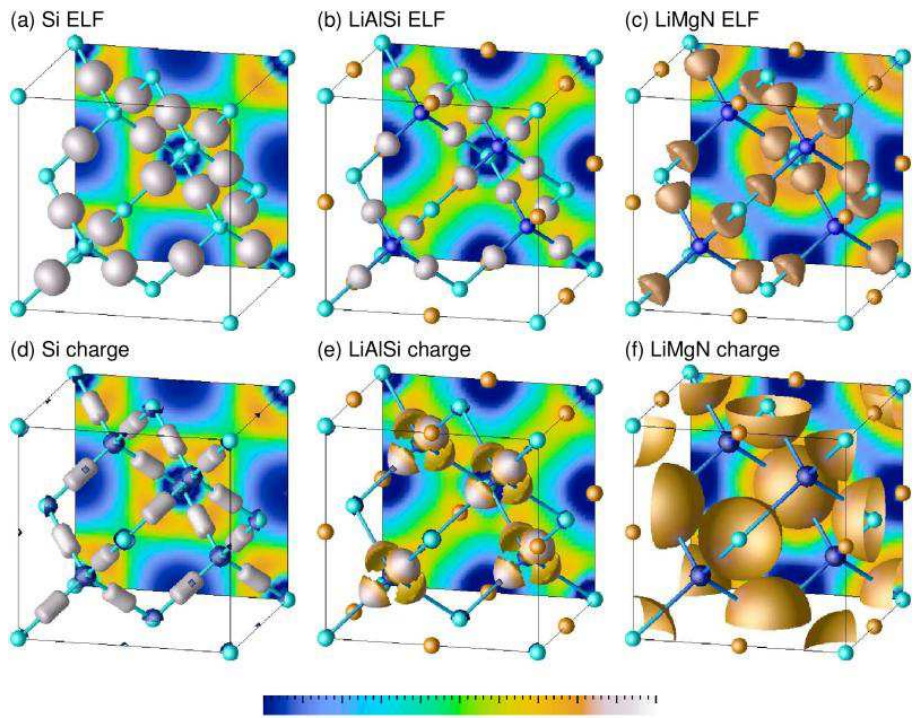}
\caption{(Color) (a), (b), and (c) are electron localization isosurfaces for 
ELF values of 0.9, 0.9, and 0.825 respectively for the three compounds 
Si, LiAlSi, and LiMgN. (d), (e), and (f) are isosurfaces of constant charge 
density at a value of 0.06\,$e$\,\AA$^{-1}$. The isosurfaces are decorated by 
the value of the electron localization function. The color-bar at the bottom of 
the figure indicates increasing localization from left (0.0) to right (1.0).
The positions of atoms are as in Figure\,1, with \textit{Z} at the origin 
\textit{etc.\/}.}
\end{figure}

In Figure\,4, we compare the electron localization function (ELF) for the three 
compounds, Si, LiAlSi, and LiMgN. Panels (a), (b), and (c) display isosurfaces
of the electron localization function for values of 0.90, 0.90, and 0.825 
respectively. These are high values of localization (the ELF scale as used
here \cite{Silvi} runs from 0 through 1) and indicate highly covalent bonding
between Si (cyan spheres) in the elemental structure, as well as between Al 
(blue) and Si (cyan) in LiAlSi, and between Mg (blue) and N (cyan) in LiMgN. 
The ELF takes on a curious hemispherical shape in LiMgN, reflecting the 
large electronegativity difference between Mg and N. The blob of localization 
is also closer to N than it is to Mg. It must be noted that there is no 
localization around Li (orange spheres) in either LiAlSi, or LiMgN, as seen 
also from the map of the ELF projected on the (010) plane at the rear 
of the unit cells. Li behaves effectively like ionic Li$^+$. In panels (d), 
(e), and (f) of this figure, we display isosurfaces of charge for a value of
0.06\,$e$\,\AA$^{-3}$ within the space of the unit cell. The charge isosurfaces
have been decorated (colored) by the ELF. Bonds distort charges from being
spherical, so distortions should be interpreted as covalency. Ionic species
on the other hand, would have spherical charge around the nucleus. We observe 
the highly covalent nature of the diamond lattice in Si, and the zinc blende 
sublattice in LiAlSi. For LiMgN, while the bonding is still covalent, the 
charge is closer to the more electronegative nitrogen. The fact that the charge
is spherical rather than directed along the bond suggests that an anionic
description might be equally valid.

\begin{figure}
\centering \includegraphics[width=7cm]{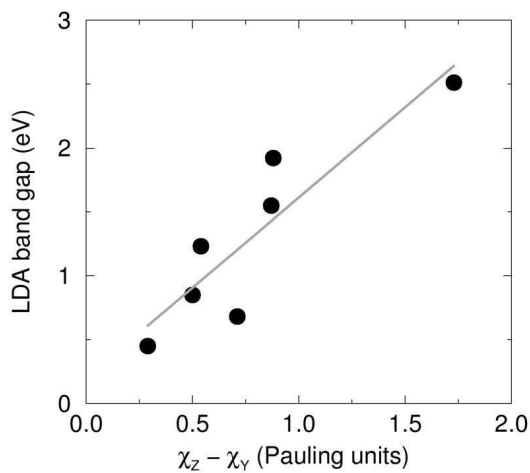}
\caption{Dependence of the LMTO energy gap of Li\textit{YZ} compounds 
on the difference in the Pauling electronegativities ($\chi$) of the \textit{Y}
and \textit{Z} species.}
\end{figure}

Having established that at least the \textit{XYZ\/} half-Heuslers with 
\textit{X} = Li can be written Li$^+$(\textit{YZ})$^-$, we use the data in
Table\,1 to examine systematics in the band gaps of these compounds. We find
the computed (LMTO) gaps of Li\textit{YZ} to vary approximately as the 
difference of the Pauling electronegativities of \textit{Y} and \textit{Z}.
The larger the difference in electronegativity of the species in the zinc-blende
sublattice (\textit{Y} and \textit{Z}), the greater the band gap as a result of
band narrowing. The trend cannot be used quantitatively, but only as
and indicator of the Phillips-van Vechten-like\cite{Phillips} 
behavior that is seen in these
complex semiconductors. It should be noted that the siting of \textit{Y}
in the tetrahedral position simultaneously allows the more polar pair of 
\textit{X} and \textit{Z} to form a stable rock salt structure. From the
viewpoint of lattice energy, this is perhaps the greater stabilizing
influence on the half-Heusler structure.

\subsection{18-electron compounds}

To examine whether similar rules hold for 18-electron half-Heusler compounds,
we start by presenting in Table\,2 the results of density functional 
calculations on a series of \textit{XYZ} compounds where \textit{X} is 
electropositive yttrium. The zinc blende lattice is formed by a later 
transition 
metal \textit{Y}, and a main group element \textit{Z}. All the compounds are 
semiconductors according to LMTO calculations, with band gaps ranging from 
0\,eV for YAuPb to 0.53\,eV for YNiAs. The existence of band gaps allows
us to formulate these phases according to the Zintl (or ``extended Zintl'')
rule X$^{3+}$(\textit{YZ})$^{3-}$ where (\textit{YZ})$^{3-}$ becomes 
isoelectronic with a diamond-structure semiconductor such as GaSb. Once again, 
we find a simple trend in the band gap with the difference in 
electronegativities of \textit{Y} and \textit{Z}, as seen from Figure\,6.

\begin{table}
\centering \begin{tabular}{|l|l|l|l|l|l|l|l|}
\hline
\multicolumn{8}{|c|}{18 electron compounds with \textit{X} = Y}\\
\hline 
\hline
\textit{XYZ} & $a_{\mathrm{Calc.}}$ (\AA) &  
       $a_{\mathrm{Exp. }}$ (\AA) & $\chi_X$ &  $\chi_Y$ 
                &  $\chi_Z$ & Gap (eV) & $B$ (GPa) \\
\hline
YNiAs   & 6.104 & 6.171 & 1.22  & 1.91 & 2.18 &  0.53 & 100.0\\  
\hline
YNiSb   & 6.350 & 6.312 & 1.22  & 1.91 & 2.05 &  0.28 & 92.8\\  
\hline
YNiBi   & 6.475 & 6.411 & 1.22  & 1.91 & 2.02 &  0.13 & 80.9\\  
\hline
YPdSb   & 6.599 & 6.527 & 1.22  & 2.20 & 2.05 &  0.16 & 92.0\\  
\hline
YAuPb   & 6.842 & 6.729 & 1.22  & 2.54 & 2.33 &  0    & 70.6\\  
\hline
\end{tabular}
\caption{Results of density functional calculations for a number of
18-electron compounds with \textit{X} = Y.}
\end{table}

\begin{figure}
\centering \includegraphics[width=7cm]{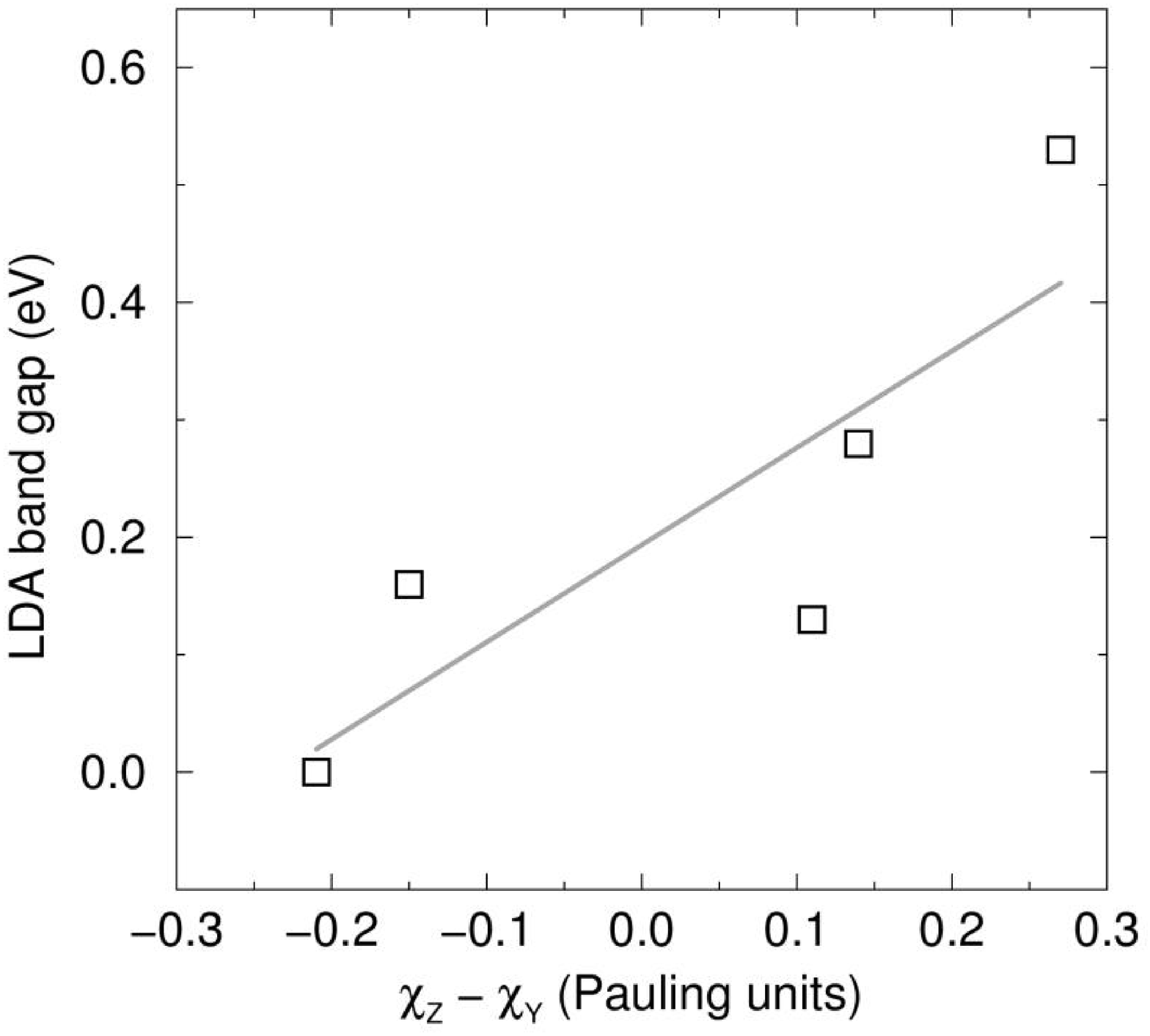}
\caption{Dependence of the LMTO energy gap of Y\textit{YZ} compounds 
on the difference in the Pauling electronegativities ($\chi$) of the \textit{Y}
and \textit{Z} species.}
\end{figure}

We proceed to examine in detail, the electronic structure of select 18-electron
half-Heusler compounds with different \textit{X}, \textit{Y}, and \textit{Z}
elements. Figure\,7(a) displays the LMTO densities of state VFeSb, TiCoSb,
and YNiSb, allowing the trends with changing the later transition metal 
\textit{Y} to emerge. Projections of the densities of state on the different
atomic levels (not displayed) reveal that the valence band has \textit{Y}
$d$ character and \textit{Z} $p$ character. The conduction band has some of the
character from these states, but in addition, has empty $d$ states from the 
\textit{X} atom. This fits with our expectation of the \textit{X} atom being
nearly fully ionized (or more accurately, having attained the group valence)
and the $d$ shell of the \textit{Y} atom being filled as a result. The gap is 
largest for TiCoSb, and smallest for YNiSb. When both \textit{X} and \textit{Y}
are changed, it is more difficult to seek trends in the gap. The gap in
VFeSb is reduced due to V and Fe not being well separated in electronegativity.
The gap in YNiSn is reduced because the unoccupied Y 4$d$ states are rather 
broad, at least within LSDA. What is evident is that compounds with Co on 
the \textit{Y} site have a strong propensity to maintain ``clean'' gaps as 
further seen in panel (b) of this figure where the DOS of TiCoSb, VCoSn, and 
NbCoSn are displayed. 

\begin{figure}
\centering \includegraphics[width=11cm]{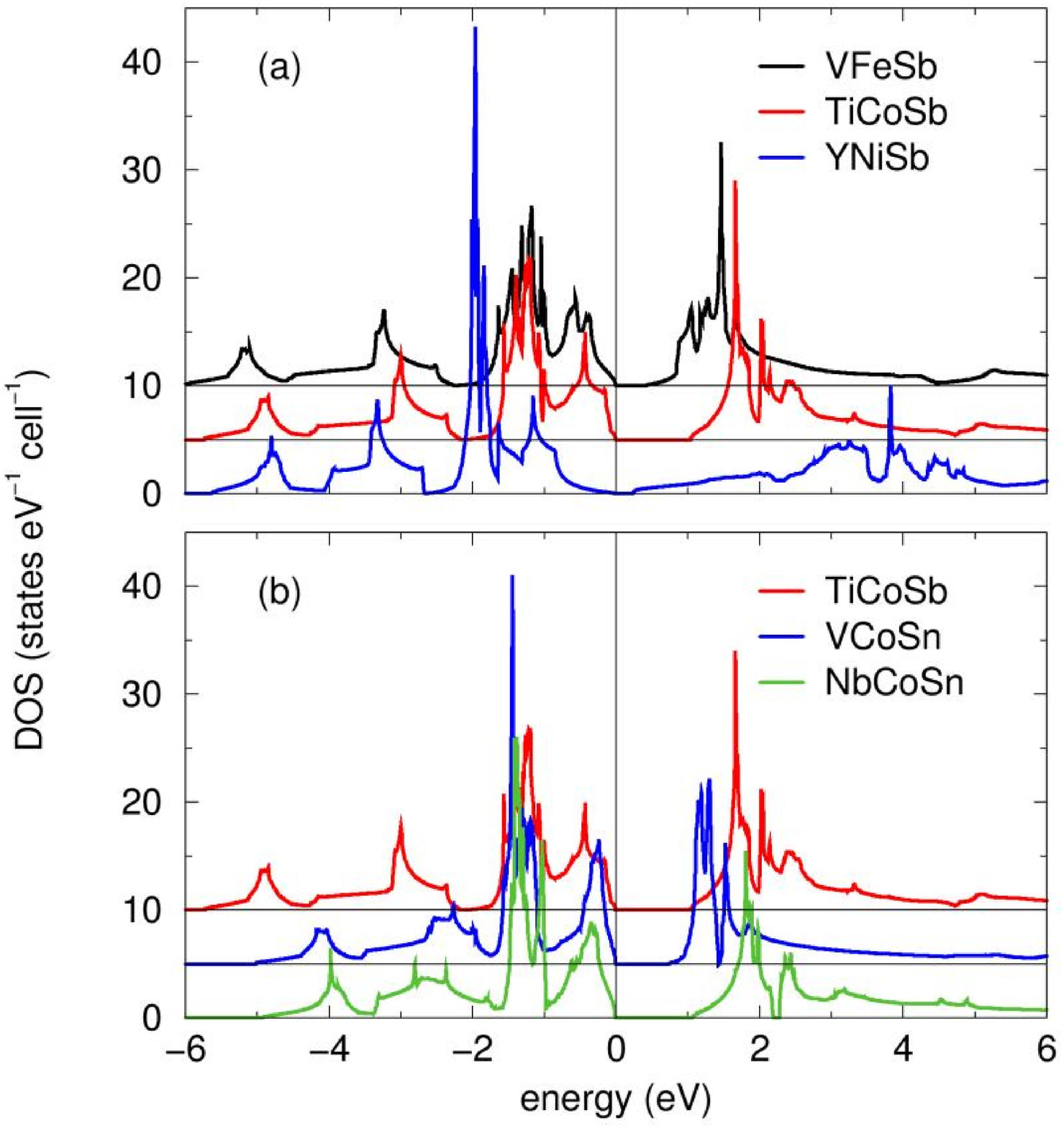}
\caption{(Color) LMTO densities of state for the 18-electron half-Heusler 
compounds (a) VFeSb, TiCoSb, and YNiSb, showing how changing the nature of the 
\textit{Y} atom affects the band gap, and for (b) TiCoSb, VCoSn, and NbCoSn
emphasizing the propensity of Co-based half-Heuslers to posses a ``clean'' gap.
The plots have been offset for clarity.}
\end{figure}

From Figure\,8, we see that even in these transition-metal rich phases,
we can perform a bonding analysis based on COHPs. These are shown for all
pairwise interactions in the two half-Heuslers VFeSb and TiCoSb. Integrating 
the COHPs, we find that the strongest bonding interactions are between 
\textit{Y} and \textit{Z} (Co and Sb, and Fe and Sb) while the interactions
between the early and late transition metal (Ti and Co, and V and Fe)
are also significant. The compounds are electronically very stable as seen 
from a complete absence of any antibonding interaction below the top of the
valence band. The fact that the 18-electron compounds are clearly valence
compounds with the band gap being located between bonding and antibonding 
levels supports our description of these phases being Zintl-like. 
The band gaps in VFeSb and TiCoSb would seem to be determined by bonding 
between \textit{X} and \textit{Y}, so at first sight, it would seem that at 
least these two transition-metal rich phases should not be described simply 
as cation-stuffed zinc blendes. However, as we shall see from the ELF analysis,
the \textit{localized\/} bonding remains in the zinc blende \textit{YZ} 
lattice. Weak bonding and antibonding COHPs between \textit{X} and \textit{Z} 
(Ti and Sb, and V and Sb) support the view that the \textit{XZ} rock salt 
sublattice is ionic in character.

\begin{figure}
\centering \includegraphics[width=11cm]{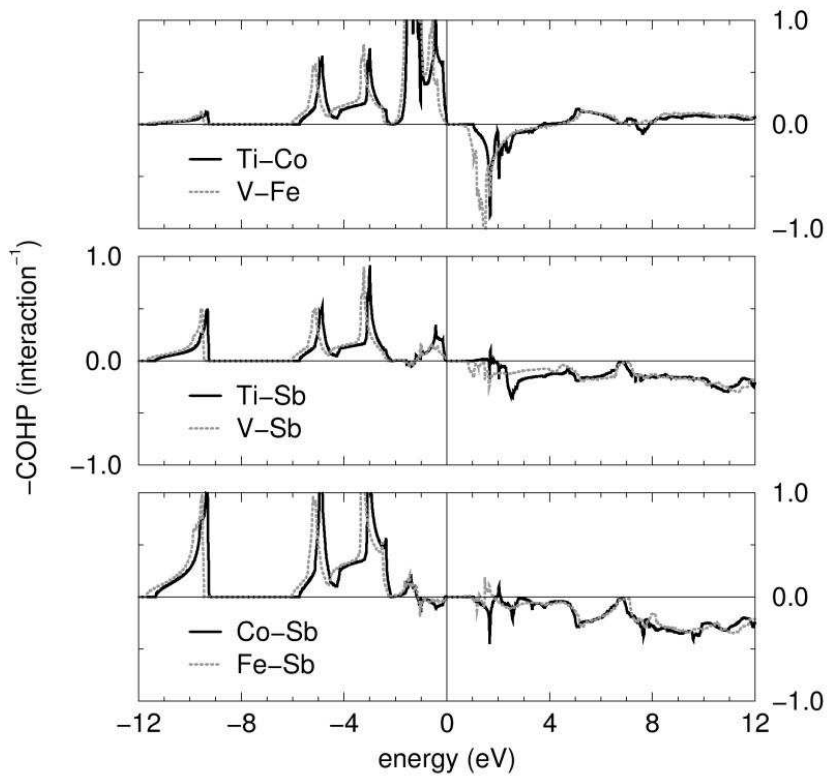}
\caption{Crystal orbital Hamiltonian populations for all pairwise 
interactions in VFeSb and TiCoSb.}
\end{figure}

Figure\,9, displays electron localization functions 
for the two 18-electron half-Heusler compounds 
for which the COHPs are displayed in Figure\,8, namely TiCoSb and VFeSb.
The ELFSs, visualized respectively for values of 0.73 and 0.71 in Figure.\,9(a)
and 9(b) are clearly indicative of strongly covalent bonding in the zinc blende
sublattice, with TiCoSb displaying the greater tendency to covalent bonding.
As we observed in the 8-electron compounds, localization isosurface is closer
to the more electronegative Sb atoms. It must be pointed out here that the
ELF is notoriously difficult to apply in $d$ electron system, and the fact 
that the localization emerges so clearly here is compelling evidence for 
dealing with these systems as if they were valence compounds with strongly
covalent bonding.

The valence charge densities displayed in Figure\,9(c) and 9(d) are very 
distinct from what was seen for the 8-electron compounds, because the filled
$d$ shell on \textit{Y} forms a large nearly spherical blobs around 
that atom, visualized for a charge density of 0.06\,$e$\,\AA$^{-3}$. We find 
that these blobs of charge are pulled out into four strongly localized (as 
seen from the coloring) lobes arranged tetrahedrally and facing \textit{Z}.
Interestingly, in VFeSb, there is also some $d$-like localization around V,
and there is some little localization seen on the isosurface between V and Fe.
The smaller electronegativity difference between V and Fe, when 
compared with the electronegativity difference between Ti and Co, in 
conjunction with the nature of COHPs displayed in Figure\,8 leads us to 
point this out as the origin for the smaller band gap of VFeSb. Another
argument that one can proffer is that TiCoSb has greater polar intermetallic
character,\cite{Brewer,Calhorda,Abdon} with strong covalent Ti-Co bonds.
V and Fe are closer together in electronegativity, and the
high formal charge state (V$^{5+}$) is much more covalent. This results
in a smaller band gap in VFeSb.

In the next subsection, we will point out that these simple ideas of covalent
bonding can be carried over to the important magnetic half-Heusler compounds.

\begin{figure}
\centering \includegraphics[width=8cm]{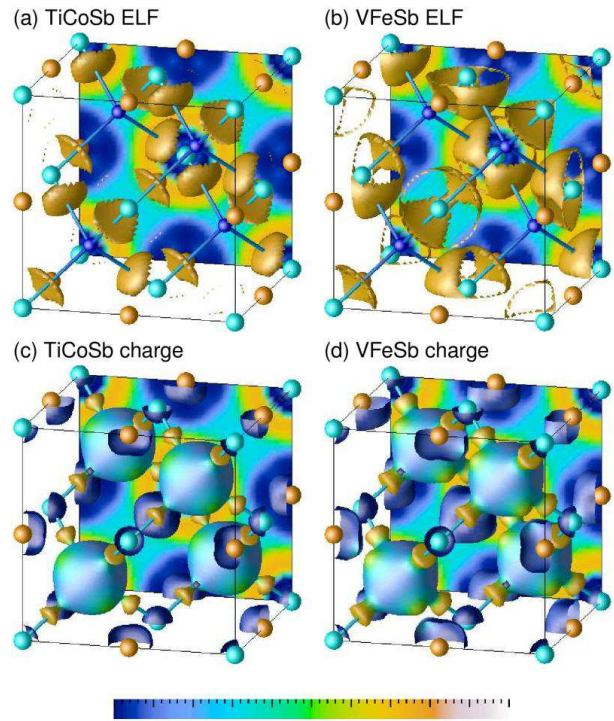}
\caption{(Color) (a) and (b) are ELF isosurfaces for TiCoSb and VFeSb for 
values of 0.73 and 0.71 respectively. (c) and (d) are charge densities
for a value of 0.06\,$e$\,\AA$^{-3}$, decorated by the ELF.}
\end{figure}

\subsection{Magnetic compounds}

\begin{table}
\centering \begin{tabular}{|l|l|l|l|l|l|}
\hline
\multicolumn{6}{|c|}{\textit{X}CoSb}\\
\hline 
\hline
\textit{X} & $a_{\rm{Calc.}}$ & $a_{\rm{Exp.}}$ & $B$ (GPa) & $n_V$ & $M$ \\ 
\hline
Sc	& 6.095 & N.A.  & 110.5	& 17 & 0 \\
\hline
Ti      & 5.888 & 5.884 & 151.5 & 18 & 0 \\
\hline
V       & 5.823 & 5.802 & 150.8 & 19 & 1 \\
\hline
Cr      & 5.820 & N.A.  & 135.4 & 20 & 2 \\
\hline
Mn      & 5.810 & 5.875 & 139.1 & 21 & 3 \\
\hline
\end{tabular}
\caption{Optimized (LAPW) and experimental cell parameters for the 
the half-Heusler compounds \textit{X}CoSb. The computed bulk moduli are also
indicated. $n_V$ is the number of valence electrons and $M$ is the 
magnetic moment obtained from LMTO calculations.} 
\end{table}

\begin{figure}
\centering \includegraphics[width=11cm]{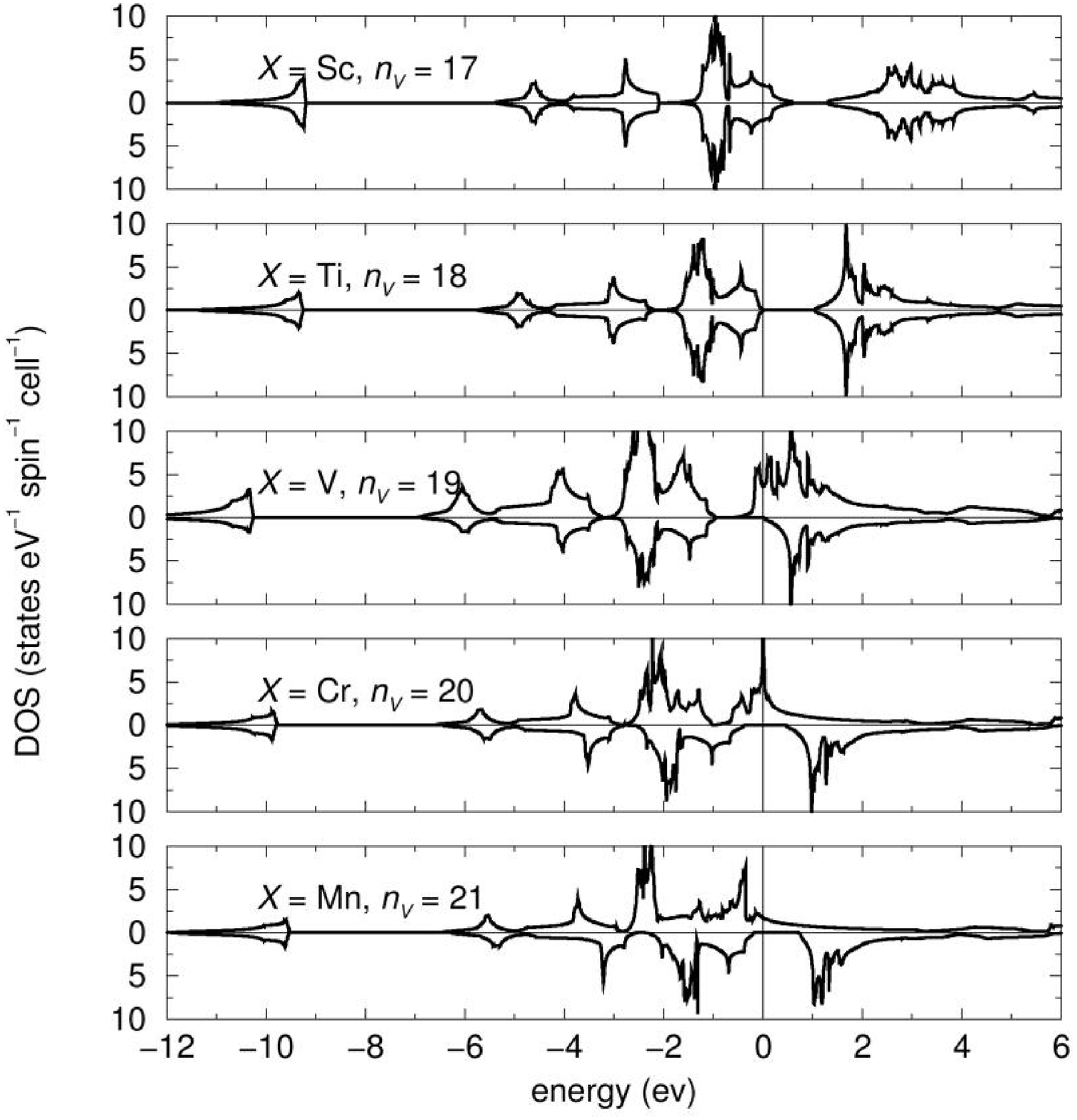}
\caption{Densities of state of the half-Heusler compounds \textit{X}CoSb, 
plotted separately in the two spin directions. The number of valence
electron ($n_V$) in each formula unit are also indicated.}
\end{figure}

The results of LAPW optimization of \textit{X}CoSb compounds are presented in 
Table\,3, with experimental cell parameters presented for comparison for the 
known compounds (\textit{X} = Ti, V, and Mn). The table also presents the 
computed (LAPW) bulk modulus, which is seen to go through a maximum for TiCoSb,
associated, as we will observe in the COHPs, with a completely filled 
\textit{bonding\/} valence band and an empty, \textit{antibonding\/}
conduction band. Figure\,10 displays densities of state for \textit{X}CoSb
phases with \textit{X} = Sc, Ti, V, Cr, and Mn. The number of valence
electrons $n_V$ per formula unit are indicated within each panel.
The DOS are plotted in the two spin directions in each panel, even for the 
non-magnetic compounds. The compound ScCoSb is not known, and neither is 
CrCoSb. The non-existence of CrCoSb could be associated with the high peak in 
the densities of state at $E_{\rm F}$ for this compound. The (hypothetical) 
compound ScCoSb is not magnetic within LMTO-LSDA because the bands are too 
broad for the Stoner criterion to be fulfilled. 

All compounds except ScCoSb obey the Slater-Pauling rules for half-Heuslers,
$M = n_V - 18$. TiCoSb with $n_V$ = 18 is a non-magnetic semiconductor,
and the calculated moments (Table\,3) for VCoSb, CrCoSb, and MnCoSb are
precisely 1, 2, and 3\,$\mu_{\rm B}$. This means that these three compounds
are half-metals, as seen from the densities of states in Figure\,10.  
In particular, CrCoSb, and MnCoSb have ``clean'' gaps in the minority spin 
direction. With increasing $n_V$, we notice that the $d$ states above the
Fermi energy, which are derived from the electropositive \textit{X} atom
drop down with respect to the filled $d$ states on Co both because of their
partial filling as well as because of the well known tendency of transition
metal $d$ levels to be stabilized in energy on going across the $d$ 
series.\cite{Jobic,Zaanen} One of the consequences of the narrowing
of the $d$ separation between \textit{Y} and \textit{X} is that for larger
valence electron counts than 21 or 22 (found when \textit{Y} = Ni),
half-Heuslers become unstable with respect to other structure types.

\begin{figure}
\centering \includegraphics[width=11cm]{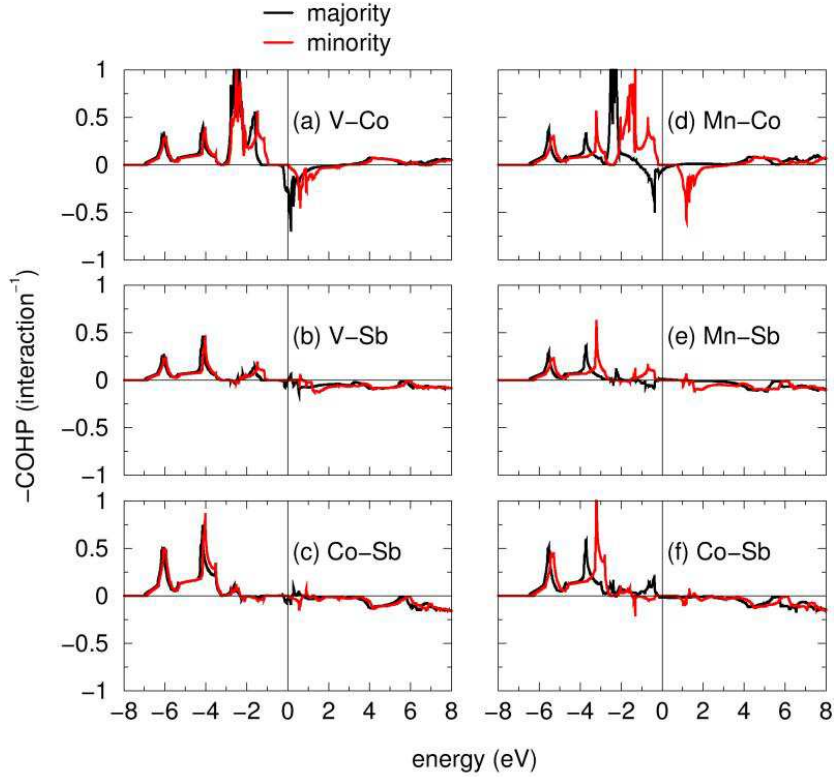}
\caption{(Color) COHPs of VCoSb and MnCoSb in the two spin directions. In the 
absence of spin-orbit coupling, majority and minority spin states do not 
interact.}
\end{figure}

In the different panels of Figure\,11, we display COHPs for 
\textit{X}-Co, \textit{X}-Sb, and Co-Sb interactions (\textit{X} = V or Mn). 
These are half-metallic ferromagnets with 19 (VCoSb) or 21 (MnCoSb)
so the COHPs are spin-resolved. While the \textit{X}-Sb interaction within the
rock-salt sublattice is clearly negligible, both \textit{X}-Co and Co-Sb are
seen to be important. The \textit{X}-Co interaction is seen to be highly 
spin-polarized. The origin of the half-metallicity is revealed by the clear
separation of antibonding \textit{majority\/} states which cross the Fermi
energy from the antibonding \textit{minority\/} states which are separated
by a gap equal to the exchange energy. This allows us to make the following
generalization: The 18-electron half-Heuslers are the most stable phases,
with well separated bonding and antibonding states. Additional electrons (more 
than 18) must go into antibonding states and these are split by 
spin-polarization and separated into majority and minority states. While 
these compounds are intrinsically less stable than the 18-electron
compounds, they maximize their stability by ensuring that minority 
antibonding states remain unoccupied. The Co-Sb interaction in these two 
compounds is seen to be strongly covalent, but not very much affected
by spin-polarization. There remains a clear separation of bonding from
antibonding states as we had observed in the 18-electron semiconductors.

\begin{figure}
\centering \includegraphics[width=12cm]{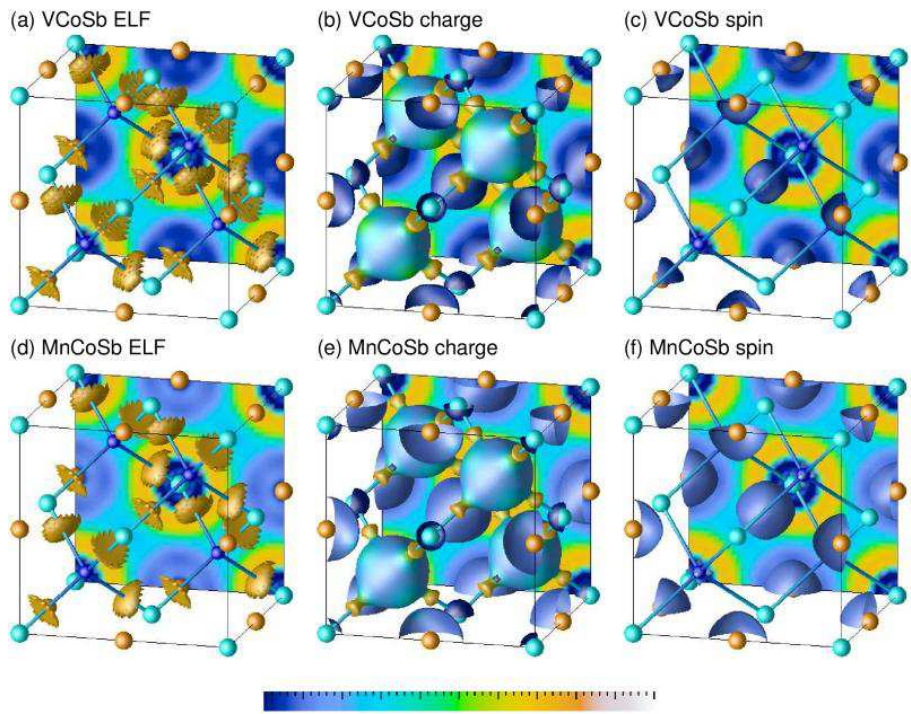}
\caption{(Color) (a) and (d) are ELF isosurfaces for VCoSb and MnCoSb for 
ELF values of 0.71. (b) and (e) are charge densities for a value of 
0.06\,$e$\,\AA$^{-3}$, decorated by the ELF. (c) and (f) are isosurfaces of
constant spin of value 0.05\,spins\,\AA$^{-3}$.}
\end{figure}

Real-space visualizations of the electronic structure in Figure\,12 reveal that
even in the magnetic compounds \textit{X}CoSb with \textit{X} = V or Mn, 
the ELFs are strongly localized on the bonds of the zinc blende CoSb network.
As could be anticipated from the similarities in the COHPs, there is almost
no change in the Co-Sb localization pattern on going from 19-electron
VCoSb to 21 electron MnCoSb [Figure\,12(a) and (d)]. There is a strongly 
localized region slightly closer to the more electronegative Sb atoms in 
both these compounds. Again, the charge density decorated by the ELF [(b) and 
(e)] confirm this localization. The $d$ electron density around Co is spherical
apart from the four lobes facing Sb. What is interesting is that the 
magnetic moment, as visualized from an isosurface of constant spin density
is clearly located on the stuffing \textit{X} atom in both in VCoSb and in 
MnCoSb, as seen in Figure\,12(c) and (f). The magnetic half-Heuslers can 
therefore be regarded as zinc-blende lattices of a late transition metal
and a main group element, stuffed by relatively electropositive magnetic ions.
Despite the presence of magnetic \textit{X}$^{n+}$ transition metal ions, 
the \textit{YZ}$^{n-}$ network can still be described in simple valence terms. 
This is reminiscent of compounds prepared by Kauzlarich and 
coworkers\cite{Kauzlarich1, Kauzlarich2} wherein magnetic ions such as 
Mn$^{3+}$ are found to behave like electropositive cations such as Al$^{3+}$
which donate charge to a closed shell  anionic sublattice. 

\subsection{MnNiSb}

\begin{figure}
\centering \includegraphics[width=11cm]{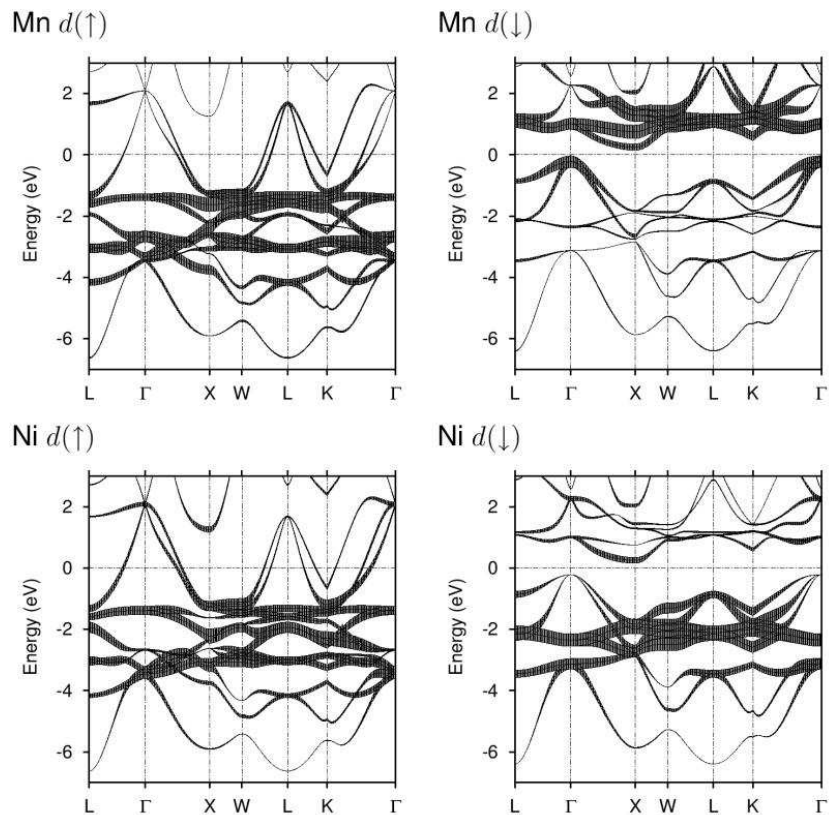}
\caption{Band structure of MnNiSb decorated by the indicated orbital 
contributions from transition metal $d$ states.}
\end{figure}

\begin{figure}
\centering \includegraphics[width=11cm]{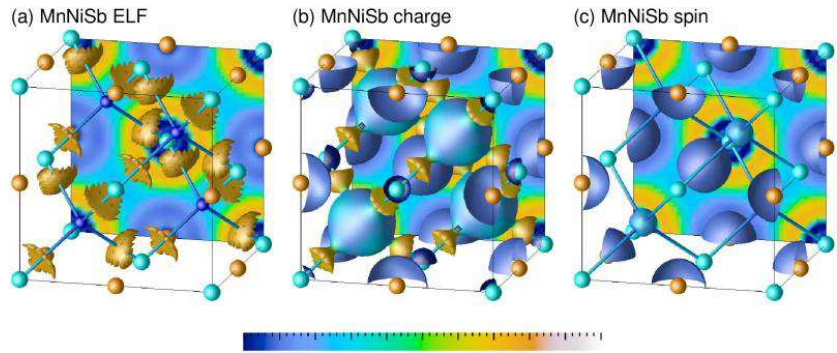}
\caption{(Color) (a) ELF isosurfaces of MnNiSb for an ELF value of 0.73. (b) 
is the charge density isosurface for a value of 0.055 $e$\,\AA$^{-3}$, 
decorated by the ELF. (c) is an isosurfaces of constant spin density 
corresponding to 0.05 spins \AA$^{-3}$.}
\end{figure}

For completion, we discuss in this subsection, the electronic structure
of the canonical 22-electron half-Heusler compound, MnNiSb.\cite{deGroot}
As we observed from the COHP of MnCoSb, we expect for MnNiSb that 
the minority gap will be formed from metal $d$ states, and this gap will fall 
within the larger gap associated with the zinc blende (NiSb) sublattice.
The panels of figure\,13(a-d) display band structure of MnNISb in the
so-called fatband representation\cite{Jepsen} where the bands are decorated 
with widths proportional to various specific orbital contributions; 
In the different panels, the $d$ orbitals of Mn and Ni are indicated in the 
different spin directions. From the band structures, it is evident that the 
magnetism resides largely on Mn. The valence band has Ni $d$ character in 
both spin direction, but only majority Mn $d$ states. The conduction band has 
minority Mn $d$ states. The disperse majority band which traverses the Fermi 
energy arises due to covalent bonding between majority Ni $d$ states and 
majority Mn $d$ states with some intermediation by Sb $p$. Indeed, we find
that he disperse band going from $\Gamma$ to L and $\Gamma$ to X is retained 
even when the calculation is performed in MnNiSb where the Sb atoms are 
replaced by empty spheres.

Figure\,14(a) and (b) display the ELF and charge of MnNiSb and confirm the 
picture of covalent bonding in the zinc blende network of this structure.
Figure\,13(c) is the spin density, and is evidently almost completely localized
on Mn in this compound as suggested by the band structure.

\ack
{Work at Mainz was supported by the \textit{Deutsche Forschungsgemeinschaft\/}
through projects FG 559 and the SPP 1166. RS gratefully acknowledges the 
National Science Foundation for support through a Career Award 
(NSF-DMR\,0449354).}

\end{document}